\documentstyle[epsfig,prl,twocolumn,aps]{revtex}
\begin{document}
\twocolumn[
\draft
\title{Boundary Effects on Spectral Properties
of Interacting Electrons in One Dimension}
\author{Sebastian Eggert, Henrik Johannesson, and Ann Mattsson}
\address{Institute of Theoretical Physics,
Chalmers University of Technology and G\"oteborg University,
S-412 96 G\"oteborg, Sweden}
\date{\today}
\maketitle
\begin{abstract}
\widetext\leftskip=0.10753\textwidth \rightskip\leftskip
The single electron Green's function of the one-dimensional Tomonaga-Luttinger 
model in the presence of open boundaries is calculated
with bosonization methods.  We show that the critical exponents
of the local spectral density 
and of the momentum distribution change in the presence of a boundary.
The well understood universal bulk behavior always crosses over to a boundary
dominated regime for small energies or small momenta.
We show this crossover explicitly for the large-U Hubbard model  
in the low-temperature limit. 
Consequences for photoemission experiments are discussed.
\end{abstract}
] \narrowtext
%\newpage
There is currently great interest in ``Luttinger liquid 
physics''\cite{Haldane,VOIT}, sparked by a new generation
of experiments on low-dimensional electron structures. Examples 
include measurements of the point contact tunneling conductance between two 
fractional quantum Hall edge channels\cite{Webb} and high-resolution 
photoemission studies of quasi one-dimensional 
metals\cite{DardelHwu,DardelVoit}. Additional interest 
stems from the fact that the Luttinger liquid - i.e. the low-energy, 
long-wavelength physics of interacting electrons in one 
dimension (1D) - provides us with the only known  
non-Fermi liquid phase with unbroken symmetry. 
For this reason the notion of a ``Luttinger liquid'' (LL) 
has played a prominent role in studies of generic features of correlated 
electron systems, and it has been suggested that some of its properties 
(e.g. anomalous propagators, spin-charge separation) carry over to higher 
dimensions\cite{Anderson}.

So far little attention has been paid to the effect of {\em boundaries}
on the spectral properties of Luttinger liquids (except in the context of
spin-chains\cite{eggert}). This is surprising 
since boundary effects are bound to be important in several of the 
proposed laboratory realizations. For example, 
in a recent series of experiments on the 
Bechgaard salts (TMTSF)$_2$X (X is a counter-ion), the HeI and II 
photoemission spectra were measured 
in the metallic phase\cite{DardelVoit}. 
This class of materials are known to exhibit strong 
electron correlations and are prime candidates for LL-behavior.  
The experimental results suggest an anomalous suppression of the  
spectral weight close to the Fermi level. It was argued in 
\onlinecite{DardelVoit} that consistency with the LL-scenario requires
long-range electron-electron interactions, but as 
we will show here, boundary effects also deplete the 
spectral weight and may influence the observed data. 
%The understanding of boundary effects is also desirable in context
%of recent  arguments \cite{Anderson} supporting the existence of 
%higher dimensional LLs which is largely based on spectroscopic data (on 
%high-$T_c$ compounds).

Specifically, we have examined the effect of an open 
boundary on the local single-particle spectral density of a spinful LL. We
have also studied the momentum distribution in a finite system with 
open boundary conditions. In both cases we observe a
strong influence from the boundaries,
which cause novel scaling behavior with energies and momenta close to
the Fermi level.

%or equivalently, for large time- and length-scales, 
%and the crossover to  ordinary ``bulk'' LL behavior may be extremely slow.
%As we shall discuss, our 
%findings have immediate consequences for the interpretation of 
%photoemission data on strongly correlated quasi-1D metals. 

As our model we take an extended version of the  
Tomonaga-Luttinger (TL) Hamiltonian, describing the 
low-energy limit of 
locally interacting 1D electrons \cite{VOIT,TomLut}, defined by the
Hamiltonian density:
%%%%%%%%%%%%%%%%%%%%%%%%%%%%
\begin{eqnarray}
{\cal H} &  = &     v_F[
  \psi^\dagger_{L,\sigma} i {d \over dx} \psi^{}_{L,\sigma} 
 - \psi^\dagger_{R,\sigma} i {d \over dx} \psi^{}_{R,\sigma}] \nonumber \\
  & + &    g_1  J_{L}^{\sigma}   J_{R}^{-\sigma}   
  \ + \    g_2  J_{L}^{\sigma}   J_{R}^{\sigma}   
  \ + \   g_3  (J_{L}^{\sigma}   J_{L}^{-\sigma} + 
J_{R}^{\sigma}   J_{R}^{-\sigma}) \nonumber \\
  & +  &   g_4  \psi^\dagger_{L,\sigma}  \psi^{}_{R,\sigma} 
  \psi^\dagger_{R,-\sigma}  \psi^{}_{L,-\sigma} 
  \label{TL}
\end{eqnarray}
%%%%%%%%%%%%%%%%%%%%%%%%%%%%%%%%%%%%%%%%
Here $J_{L/R}^\sigma
 \equiv :\psi_{L/R,\sigma}^\dagger\psi_{L/R,\sigma}^{}:$ are the 
chiral Fermion currents of the left and right moving components
of the electron field $\Psi_{\sigma}(x)$, expanded about the
Fermi points $\pm k_F$: $\Psi_{\sigma}(x) = %\frac{1}{\sqrt{2}}
e^{-ik_Fx}\psi_{L,\sigma}(x) + e^{ik_Fx}\psi_{R,\sigma}(x).$ 
The Hamiltonian (\ref{TL}) describes left- and
right moving relativistic Fermions in (1+1)-dimensions which interact
via forward scattering without spin flip ($g_1, g_2,$ and $g_3$) or 
with spin flip ($g_4$). ``Umklapp'' processes are suppressed 
away from half-band-filling, 
so this model provides a complete picture of possible local interactions
in this case.    
(Trivial forward scattering terms which can be absorbed by redefining
the Fermi velocity have not been explicitly written out.) 

The TL Hamiltonian is conveniently bosonized \cite{Stone} by introducing 
charge and spin currents and the corresponding
bosons $\phi_c$ and $\phi_s$ with 
conjugate momenta $\Pi_c$ and $\Pi_s$, respectively:
\begin{equation} 
J_L^{c/s} \equiv \frac{1}{\sqrt{2}} \left( J_L^\uparrow \pm
J_L^\downarrow \right) = \frac{1}{\sqrt{4 \pi}} \left(\Pi_{c/s} + 
{\partial_x \phi_{c/s}}\right)
\end{equation}
and accordingly for right-movers.
The resulting theory describes separate spin and charge
excitations moving with velocities 
$v_c = { v_F \over 2} + {g_3 \over 2 \pi}$ and
$v_s = { v_F \over 2} - {g_3 \over 2 \pi}$, respectively:
\begin{eqnarray}
{\cal H}  & =  & \sum_{\nu=s,c}\left\{    v_\nu
\left[({\partial_x \phi_\nu})^2 + \Pi_\nu^2\right]
\ + \ \frac{g_\nu}{4 \pi} 
\left[({\partial_x \phi_\nu})^2 - \Pi_\nu^2\right] \right\}
\nonumber \\
& + & g_4 \ {\rm const.} \cos \sqrt{8 \pi} \phi_s \label{boson-ham},
\end{eqnarray}
where
$g_c = g_1 + g_2$ and $g_s = g_2 - g_1$.
The charge interaction $g_c$
can be absorbed into the free Hamiltonian by a
simple rescaling of the charge boson, but the  spin
interactions  $g_s$ and $g_4$ obey Kosterlitz-Thouless renormalization
group equations \cite{Kosterlitz} 
with flow lines along hyperbolas $g_s^2-g_4^2 = \rm const.$
(to lowest order).  For $g_s > -|g_4|$ the spin sector develops a gap in the
low energy, long wave-length limit, but for $g_s < -|g_4|$ the system flows
to a stable fixed point $g_s^* = -\sqrt{g_s^2 -g_4^2}, \ g_4^* = 0$.  For
$g_s = -|g_4|$ the interaction corresponds to one single marginally
irrelevant operator, so that $g_s^*=g_4^*=0$.
If the flow to a stable fixed point occurs, we can rescale the bosons 
by a canonical transformation to obtain a free theory ($\nu= s,c$)
% but to understand the  spin
%interactions we have to consider the corresponding renormalization 
%group equations for $g_s$ and $g_4$
%by lowering the infrared cut-off $\Lambda$ (i.e. the finite temperature
%or inverse length).  To lowest order
%we have $\partial g_4 /\partial \ln \Lambda = g_s g_4,  
%\ \partial g_s /\partial \ln \Lambda = g_4^2$, 
%so that the flow-lines are along hyperbolas.  
\begin{equation}
\phi_\nu \to {K_\nu \phi_\nu}, \ \ \ \Pi_\nu \to {\Pi_\nu / K_\nu},
\end{equation}
where to first order in the coupling constants
\begin{equation} 
K_s^2 = 1 - {g_s^*}/{4 \pi v_s}, \ \ \ K_c^2 =1-{g_c}/{4 \pi v_c}.
\end{equation}

We now consider a semi-infinite system with  an 
{\it open} boundary condition at the origin and thus require the electron 
field $\Psi_{\sigma}(x)$ to vanish at $x=0$. This  implies  
$\psi_{L,\sigma}(0) = - \psi_{R,\sigma}(0)$, or in terms of the bosons
$\phi_{L,c}(0) =  -\phi_{R,c}(0) + \sqrt{\pi}/{2K_c}$ and 
$\phi_{L,s}(0) =  -\phi_{R,s}(0)$, which
allows an analytic continuation of the left-movers onto the
negative half axis in terms of right-movers 
%%%%%%%%%%%%%%%%%%%%%%%%%%%%%%%%%%%%%%%%%%%%%%%%%%%%%%%%%
\begin{equation}
\phi_{L,\nu}(x,t) =  -\phi_{R,\nu}(-x,t) + {\rm const.}, \ \ \ \ 
 x < 0 \label{openBC}
\end{equation}
($\nu = s,c, \ \ {\rm const.} = 0,\sqrt{\pi}/{2K_c}$).
%%%%%%%%%%%%%%%%%%%%%%%%%%%%%%%%%%%%%%%%%%%%%%%%%%%%%%%%%%%%%%
We can therefore describe the theory in terms of left-movers only
which live on the full complex plane {\it without} an
explicit  boundary condition.  Using this formalism, the single electron
Green's function can be calculated in a straightforward way:
\begin{eqnarray}
&&  <\Psi^{\dagger}_\sigma(x,t) \Psi^{}_\sigma(y,0) >   \nonumber \\
&= &  e^{i k_F (x-y)}G(x,y,t) \ + \ e^{-i k_F (x-y)}G(-x,-y,t) \nonumber \\
&  -&   e^{i k_F (x+y)}G(x,-y,t) \ - \
e^{-i k_F (x+y)}G(-x,y,t),  \label{fullGF}
\end{eqnarray}
where the chiral Green's function $G(x,y,t)$ 
is a product of spin and charge contributions
\begin{eqnarray}
G(x,y,t) \ \propto & &
\prod_{\nu = c,s}
\left({v_\nu t+ x-y}\right)^{-\frac{(K_\nu+K_\nu^{-1})^2}{8}} 
\nonumber \\ & \times &
\left({v_\nu t- x+y}\right)^{-\frac{(K_\nu-K_\nu^{-1})^2}{8}} 
\nonumber \\ & \times &
\left( \frac{|4 x y|} {[v_\nu^2 t^2- (x+y)^2]}
\right)^\frac{K_\nu^{-2}-K_\nu^{2}}{8}
 \label{GF}
\end{eqnarray}
Here, $x$ and $y$ denote the distance from the boundary ($x=0$) and the
time carries an implicit ultraviolet cut-off $t - i \epsilon$.
We can see that in the limit $xy \gg |(x-y)^2 - v_\nu^2 t^2|$ 
the last factor in equation (\ref{GF})
goes to unity and we recover the known bulk correlation function\cite{VOIT}
(as we do in the non-interacting case $K_c = K_s = 1$).
 
To understand the physical implications of the boundary correlation
function we study the {\it local} spectral density $N(\omega,r)$,
which is given in terms of the Green's function as
\begin{equation}
N(\omega,r)  \equiv  \frac{1}{2 \pi} \int_{-\infty}^\infty
\!\!e^{i \omega t} \langle\left\{
\Psi^{\dagger}_\sigma(r,0), \Psi^{}_\sigma(r,t)\right\}\rangle dt,
\label{N_omega}
\end{equation}
where $\omega$ is measured relative to the Fermi energy and
$r$ is the distance from the boundary.
Without the boundary, this integral can be done exactly\cite{MedenSchonhammer}
with the result that the spectral density scales at the
Fermi surface as $N(\omega) \propto \omega^\alpha$,
where the exponent in the bulk is given by 
\begin{equation}
\alpha = (K_c^2 + K_c^{-2} + K_s^2 + K_s^{-2})/4 -1. \label{alpha}
\end{equation}

However, the boundary clearly has an effect on this exponent,
and simple power counting shows that we expect a  crossover 
to a boundary dominated regime for $\omega < v_\nu/r$ with a novel
exponent $\alpha_B = (K_c^{-2} + K_s^{-2})/2 -1$. 
Interestingly, the boundary exponent $\alpha_B$ therefore {\it always}
dominates for sufficiently small $\omega$.  Moreover, we notice that
the last two terms in equation (\ref{fullGF}) make a contribution which
oscillates at twice the Fermi wave-vector and drops off with
the distance from the boundary proportional to
$e^{i 2 k_F r} r^{-(K_c^2 +K_s^2)/2}$.  This contribution
is reminiscent of a Friedel oscillation, although it can 
probably not be observed directly, since
experimental measurements of the density of states (e.g. photoemission)
will average over several lattice sites.  We therefore ignore those 
``Friedel'' terms in the following calculations. 
 
\begin{figure}
\begin{center}
\mbox{\epsfig{file=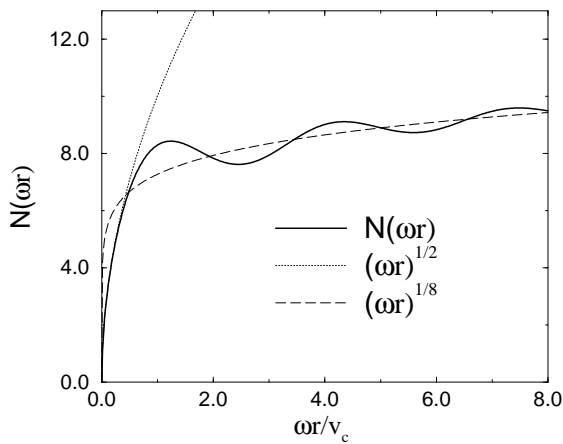,width=3.35in,angle=270}}
\end{center}
\caption{The spectral density as a function of $r\omega$ in arbitrary units.
The corresponding 
power-laws for $\alpha = 1/8$ and $\alpha_B= 1/2$ are also shown.} 
\label{n_ro}
\end{figure}

As an example, we consider
the low temperature Hubbard model away from half-filling
 which is well understood in terms
of the TL model\cite{VOIT}. 
In this case, the SU(2)-invariance forces $K_s =1$, and
it is known from Bethe ansatz calculations that $K_c^2 \to 1/2$ as
$U \to \infty$\cite{LiebWu}. From these numbers, 
the well known result $N(\omega) \propto |\omega|^{1/8}$
for small $\omega$ follows immediately for the bulk regime 
$\omega \gg v_c/ r$.   
In the presence of the boundary, however, we crossover to the boundary
exponent $\alpha_B = 1/2$ for $\omega < v_c/r$.  
After rescaling the variable of integration
in equation (\ref{N_omega}), we see that the spectral density  is a
function of the scaling variable $r \omega$ only (up to an overall constant):
$N(r, \omega) = r^{-1/8} \ f(r \omega)$.  
After a deformation
of the integration contour, a numerical integration of equation (\ref{N_omega})
is straightforward (we subtract the divergent part).
The results
in figure (\ref{n_ro}) clearly show the crossover from boundary behavior 
for $r \omega/v_c < 1$ with exponent $\alpha_B=1/2$ 
to bulk behavior for $r \omega/v_c > 1$ with
exponent $\alpha=1/8$ (the corresponding
power-laws are superimposed). The observed oscillations in figure 
(\ref{n_ro}) are an interesting secondary effect, which
vanish asymptotically as $\sin (2 \omega r/v_c) (\omega r)^{-13/16}$,
(but those are {\it not} due to the ``Friedel'' 
terms in equation (\ref{fullGF}) which have been neglected).

Our results have direct relevance for photoemission experiments on 
quasi-1D metals. 
%The experimental resolution in these experiments
%is about $\Delta \approx 20meV$ which is two orders of magnitude 
%lower than the Fermi energy, so that the condition for boundary behavior
%$r \omega/v_c < 1$ can be fulfilled for distances $r$ of about
%one-hundred lattice spacings $a$ (since $v_c \sim a E_F$).  
%This means that boundary effects are
%always observed for frequencies close to the Fermi energy
%in materials with about 1\% impurities or more.
%Another way of probing the boundary regime is to use single crystals
%and measure close to the edges (e.g. probing electrons which escape 
%from the crystal face where the open ends of the molecular chains are).
At low temperatures the photoemission intensity 
$I(\em \omega)$ is proportional to the local spectral density 
$N (\omega , r)$, integrated over the region  of escaping electrons
 and weighted by the Fermi-Dirac distribution $f_{FD}(\omega)$:
\begin{equation}
I(\omega) \propto \int dr \ f_{FD}(\omega) N(\omega , r) \ . 
\label{intensity}
\end{equation}
(We neglect the small thermal shift in $N(\omega , r)$ at 
low temperatures.)

\begin{figure}
\begin{center}
\mbox{\epsfig{file=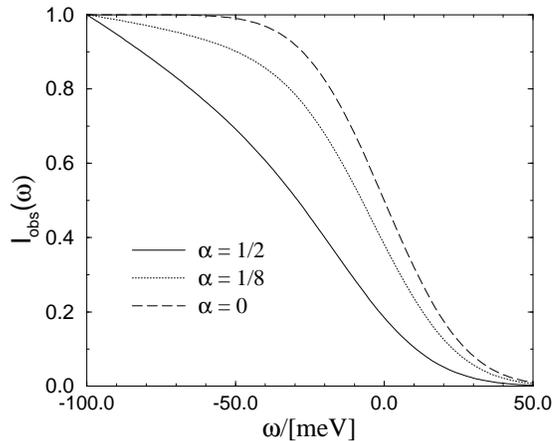,width=3.35in,angle=270}}
\end{center}
\caption{The predicted intensity $I_{\rm obs}$ in arbitrary units as a
function of $\omega$
for boundary and bulk cases (i.e. for power-laws with $\alpha_B=1/2$ and
$\alpha = 1/8$, respectively). The corresponding three-dimensional case
($\alpha = 0$) is also shown.
Temperature ($T=50K$) and finite resolution ($\Delta = 20meV$) effects have
been taken into account according to equations 
(\protect{\ref{intensity}}) and (\protect{\ref{GaussAverage}}). }
\label{avg}
\end{figure}

In a boundary dominated region,  
$I(\omega)$ is seen to be dramatically reduced at the Fermi level 
compared to a ``bulk region''.
Moreover, the finite energy resolution $\Delta$
of the photon lines effectively 
introduces an averaging over the ``true'' spectral density
\begin{equation}
I(\omega)_{\rm obs} \equiv \frac{1}{\sqrt{2 \pi} \Delta}
\int e^{-(\omega-x)^2/2 \Delta^2} I(x) \ dx.  \label{GaussAverage}
\end{equation}
This averaging completely wipes out the power-law singularities in either
the bulk or the boundary case as shown in figure (\ref{avg}), where
we plotted $I(\omega)_{\rm obs}$ in arbitrary units 
at $T=50K$ for boundary ($\alpha_B = 1/2$)
and bulk ($\alpha = 1/8$) regimes, respectively, 
assuming an experimental resolution of $\Delta = 20 meV$ 
(experimental values according to \onlinecite{DardelVoit}).
The corresponding three-dimensional case ($\alpha=0$) is also shown for
comparison.  In the neighborhood of the Fermi level the observed 
boundary dominated spectral density appears to be depleted with an
exponent of one or larger
(compared to the exponent $\alpha = 1/8$ of the bulk spectral function
without temperature or averaging effects). 

In experiments the condition for boundary behavior $\omega < v_c/r$
will be fulfilled over an energy range $\omega \sim E_F/\ell$, where $\ell$
is the distance from the boundary in units of the lattice spacing. 
This means that if the effective length $\ell$ of the 1D chains
is about one-hundred lattice spacings, the
boundary effects are observed over a region of $\alt 100 meV$
around the Fermi energy (e.g. corresponding to an ``impurity'' 
density of 1\%  close to the cleaved surface).
Turning to the photoemission data reported in \cite{DardelVoit},    
a combination of boundary effects and a finite 
experimental resolution go a long way to account for the observed  
suppression of spectral weight at the Fermi level. 
The experiments indeed suggest a scaling $I(\omega)_{\rm obs} \propto
\omega^{\alpha_{\rm obs}}$, with $\alpha_{obs} > 1$ extending, however,
over a larger energy interval, and some 
additional mechanism (long-range 
interactions\cite{DardelVoit,MedenSchonhammer} or
electron-phonon coupling \cite{VoitSchulz})
may have to be invoked to fully explain the data. 
Another way of directly examining the boundary region would be to probe
the single crystal face where the open ends of the chains are.
Since the electrons have an escape depth of only a few lattice spacings, the
boundary exponent should be observable over a much larger energy range.

As a second example of the effect of open boundaries, 
we consider the momentum distribution of a {\it finite}
system with length $L$ and open boundary conditions, which can be
expressed in terms of the chiral Green's function
\begin{equation}
n(k,L) \equiv  \frac{1}{L}   \int_0^L \! dx \  dy \  
\cos k(x-y) \ G(x,y,0),
\label{n_k} \end{equation}
where $k$ is measured relative to the Fermi wave-vector $k_F$
(we also implicitly subtract $n(0,L)$ to remove any divergences).
The bulk behavior can be determined by taking 
$L \to \infty$ and by simple power-counting we see that
$n(k, \infty) \propto |k|^\alpha$, where $\alpha$ is
again given in equation (\ref{alpha}).
However, boundary effects will be present,
and moreover we have to consider that 
we are now dealing with a {\it finite} 
system (i.e. open boundary conditions both
at $x=0$ and $x=L$).  One way of determining the correct
correlation functions in this case is to assume conformal invariance 
in the complex plane $z' = x + iv_\nu \tau$ 
(which is justified in the low energy, 
long-wavelength limit and with decoupled spin and charge sectors).
We then simply perform a conformal transformation onto a
cylinder with circumference $2L$: $z' \to e^{i  \pi z/L}$ [after the
analytic continuation in equation (\ref{openBC})].
Since we know the chiral Green's function in the plane ($z'$), we immediately
obtain the result for the finite case\cite{fabrizio}
(using the transformation rules for primary fields)
\begin{eqnarray}
G(x,y,t) & \propto & 
\prod_{\nu = c,s}
\left({\frac{2L}{\pi}\sin\frac{\pi(v_\nu t+ x-y)}{2L}}
\right)^{-\frac{(K_\nu+K_\nu^{-1})^2}{8}} 
\nonumber  \\ & \times &
\left({\frac{2L}{\pi}\sin\frac{\pi(v_\nu t- x+y)}{2L}}
\right)^{-\frac{(K_\nu-K_\nu^{-1})^2}{8}} 
 \label{finiteGF} \\ & \times &
\left( \frac{\sin\frac{\pi x}{L}\sin\frac{\pi y}{L} } 
{\sin\frac{\pi(v_\nu t+ x+y)}{2L}
\sin\frac{\pi(v_\nu t- x-y)}{2L}}
\right)^\frac{K_\nu^{-2}-K_\nu^{2}}{8}. \nonumber
\end{eqnarray}

We expect the critical behavior to be changed dramatically by this
transformation, but  
now the exponent cannot be determined by simply counting the
powers in the finite Fourier transform.  Again, the momentum
distribution is a function of a scaling variable $kL$ only (up to 
an overall constant): $n(k,L) = L^{-\alpha} f(kL)$.  We are
mostly interested in the {\it apparent} 
critical exponent $\alpha_{\rm app}$ near the Fermi wave-vector
$n(k, L) \propto |k|^{\alpha_{\rm app}}$. This apparent exponent will
change slowly depending on the scale $kL$ at which we probe the system,
so it is useful to define a scale dependent exponent in terms of the
logarithmic derivative
\begin{equation} 
\alpha_{\rm app}(kL)  \equiv  
\frac{k}{n(k,L)} \frac{\partial n}{\partial k}(k,L). \label{log-deriv}
\end{equation}

\begin{figure}
\begin{center}
\mbox{\epsfig{file=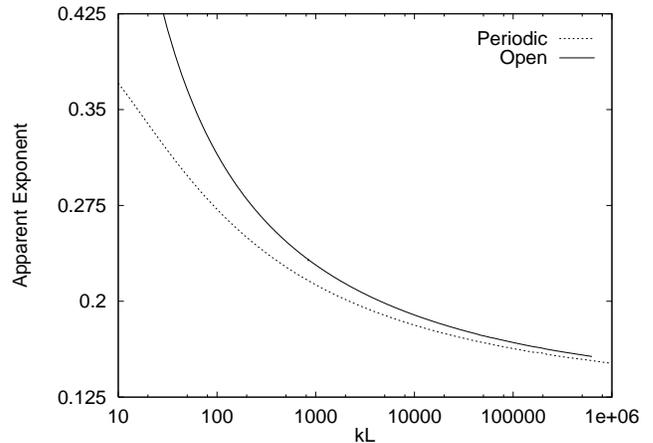,width=3.35in,angle=0}}
\end{center}
\caption{The apparent exponent $\alpha_{\rm app}$ 
of the momentum distribution for a finite system as a function of $kL$
according to equation (\protect{\ref{n_k}}). As $kL \to \infty$, the
bulk exponent $\alpha = 0.125$ is slowly approached.} \label{alpha_app}
\end{figure}

As our example we consider the case of the large-$U$ Hubbard model again. 
After calculating
$\alpha_{\rm app}$ numerically as shown in figure (\ref{alpha_app}), we  find
that the crossover is extremely slow.  Even for huge values of $kL \sim 10^{6}$
we are still considerably far away from the accepted bulk exponent 
$\alpha = 0.125$.  Since $k$ is assumed to be small compared to $k_F$, 
this means that even macroscopic samples of several centimeters will have
a finite-size dominated momentum distribution near $k_F$.  
A slow crossover is also observed for a {\it periodic} (but finite)
system, which is shown for comparison in figure  (\ref{alpha_app}).
The observed behavior is therefore mostly a finite size effect, but
systems with an {\it open} boundary condition show an even slower crossover.
%Interestingly, the apparent critical exponent $\alpha_{\rm app}$
%of the momentum distribution is {\it not} the same as the one observed in the
% spectral density $\alpha_{\rm obs}$ (unlike in 
%``bulk'' systems, where the same exponent for both quantities
%is considered a special feature of the LL).

In conclusion we have shown that an open boundary has a
pronounced influence on the observed critical exponents in quasi 
one-dimensional metals.  The
spectral density was shown to be {\it always} dominated by the boundary
exponent for frequencies close to the Fermi energy, which has
direct consequences for the interpretation of photoemission experiments. 
The momentum distribution exhibits strong finite size and also boundary 
effects close to the
Fermi wave-vector and the crossover to ``bulk'' behavior is extremely slow.

\begin{acknowledgements}
We thank I. Affleck, A.A. Nersesyan, and K. Sch\"onhammer for valuable 
    discussions, and M. Grioni for  correspondence about 
    photoemission experiments. This research was supported in part by
    the Swedish Natural Science Research Council.
\end{acknowledgements} 
%%%%%%%%%%%%%%%%%%%%%%%%%%%%%%%%%%%%%%%%%%%%%%%%%%%%%%%%%%%%


\begin{thebibliography}{19}
\bibitem{Haldane} F. D. M. Haldane, {\it J. Phys. C: Solid State Phys.}
{\bf 14}, 2585 (1981).
\bibitem{VOIT} For a recent review, see J. Voit, ``One-Dimensional 
Fermi Liquids'', {\tt cond-mat/9510014}, to appear in {\it
Rep. Prog. Phys.}
\bibitem{Webb}  F. P. Milliken, C. P. Umbach, R. A. Webb, to
appear in {\it Solid State Comm.}
\bibitem{DardelHwu} B. Dardel, D. Malterre, M. Grioni, 
P. Weibel, Y. Baer, H. L\'evy, {\it Phys. Rev. Lett.}
{\bf 67}, 3144 (1991); Y. Hwu, P. Alm\'eras, M. Marsi, H. Berger, F. L\'evy, 
M. Grioni, D. Malterre, G. Margaritondo, 
{\it Phys. Rev.} {\bf B46}, 13624 (1992).
\bibitem{DardelVoit} B. Dardel, D. Malterre, M. Grioni, P. Weibel, Y. 
Baer, J. Voit,  D. J\'er\^ome, {\it Europhys. Lett.} {\bf 24}, 687 (1993). 
\bibitem{Anderson} P.W. Anderson, {\it  Phys. Rev. Lett.} {\bf 64}, 1839 
(1990).
\bibitem{eggert} S. Eggert, I. Affleck, {\it Phys. Rev.} {\bf B46}, 10866 
(1992) and {\it Phys. Rev. Lett.} {\bf 75}, 934 (1995). 
\bibitem{TomLut} S. Tomonaga, {\it Prog. Theor. Phys.} {\bf 5}, 544
(1950); J. M. Luttinger, {\it J. Math. Phys.} {\bf 4}, 1154 (1963).
\bibitem{Stone} For a collection of relevant papers, see M. Stone (ed.), 
{\em Bosonization}, (World Scientific, Singapore, 1994). 
\bibitem{Kosterlitz} 
 J.M. Kosterlitz, {\it J. Phys. C: Solid State Phys.} {\bf 7}, 1046 (1974).
\bibitem{MedenSchonhammer} K. Sch\"onhammer,  V. Meden, {\it Phys. Rev.} 
{\bf B47}, 16205 (1993) and (E) {\bf B48}, 11521 (1993); J. Voit,
{\it J. Phys. C: Cond. Matt.} {\bf 5}, 8305 (1993).
\bibitem{LiebWu}  E.H. Lieb, F.Y. Wu, {\it Phys. Rev. Lett.} {\bf 20},
1445 (1968).
\bibitem{VoitSchulz} J. Voit, H. J. Schulz, {\it Phys. Rev.} {\bf B37}, 
10068 (1988).
\bibitem{fabrizio}  Equation (\ref{finiteGF})  agrees with 
an independent result by M. Fabrizio, A.O. Gogolin, 
 {\it Phys. Rev.} {\bf B51}, 17827 (1995),
where a finite-mode expansion of the boson was used instead.
\end{thebibliography}
\end{document}